\documentclass{article}

\usepackage{ismir,amsmath,cite}
\usepackage[bookmarks=false]{hyperref}
\usepackage{graphicx}
\usepackage{color}

\usepackage{tikz}
\usepackage{amssymb,mathtools}
\usepackage{breqn}
\usepackage[norelsize,linesnumbered]{algorithm2e}
\newcommand{\smallurl}[1]{{\tiny\url{#1}}}

\usepackage{booktabs}
\newcommand{\vect}[1]{\boldsymbol{ #1}}

\title{From Bach to the Beatles: The simulation of human tonal expectation using ecologically-trained predictive models}

\multauthor
{Carlos Cancino-Chac\'on$^{1,2}$ \hspace{1cm} Maarten Grachten$^2$ \hspace{1cm} Kat Agres$^3$} 
{ \bfseries{\hspace{1cm}}\\
 $^1$ Austrian Research Institute for Artificial Intelligence (OFAI), Vienna, Austria\\
$^2$ Department of Computational Perception, Johannes Kepler University, Linz, Austria\\
$^3$ Institute of High Performance Computing, A*STAR, Singapore\\
{\tt\small carlos.cancino@ofai.at, maarten.gracthen@ofai.at, kat\_agres@ihpc.a-star.edu.sg}
}
\def\authorname{Carlos Cancino-Chac\'on, Maarten Grachten, Kat Agres}

\begin{document}

\maketitle
%

\begin{abstract}
Tonal structure is in part conveyed by statistical regularities between musical events, and research has shown that computational models reflect tonal structure in music by capturing these regularities in schematic constructs like pitch histograms. Of the few studies that model the acquisition of perceptual learning from musical data, most have employed self-organizing models that learn a topology of static descriptions of musical contexts. Also, the stimuli used to train these models are often symbolic rather than acoustically faithful representations of musical material.
In this work we investigate whether sequential predictive models of musical memory (specifically, recurrent neural networks), trained on audio from commercial CD recordings, induce tonal knowledge in a similar manner to listeners (as shown in behavioral studies in music perception). Our experiments indicate that various types of recurrent neural networks produce musical expectations that clearly convey tonal structure. Furthermore, the results imply that although implicit knowledge of tonal structure is a necessary condition for accurate musical expectation, the most accurate predictive models also use other cues beyond the tonal structure of the musical context.
\end{abstract}



\section{Introduction and related work}\label{sec:introduction}


Computers are increasingly being used to perform music-related tasks (automated music analysis, music recommendation, composition, etc).
To perform such tasks reliably, there is a need for computers to grasp concepts that are relevant to our perception and understanding of music~\cite{DBLP:journals/tist/Widmer17}.
Empirical findings from music psychology are valuable in this respect, since they shed light on the process of human music perception and cognition.

We know from extensive research in music psychology that listeners implicitly extract statistical properties governing tonal structure through exposure to music \cite{krumhansl1990,saffran1999statistical,agres2017information}. The tonal \textit{stability}, or relative importance, of notes in a key may be largely due to the frequency of occurrence of pitches in a piece of music. 
The more foundational pitches (e.g., C, E, and G in the key of C major) will tend to be anchor points in the music, and will often occur on metrically-important positions \cite{Krumhansl:2010ji,Lerdahl:1983ti}.

Through exposure to these kinds of melodic (and harmonic) statistical properties, listeners form an implicit mental model of tonality. Evidence for this has been provided, for example, through the seminal work of Krumhansl and colleagues employing a `probe-tone paradigm', in which listeners rate how well the last pitch, or probe-tone, of a musical sequence fits in with the previous context.
When provided with a tonal context, such as an ascending or descending musical scale, listeners perceive certain pitches as sounding more appropriate than others \cite{krumhansl1990,Krumhansl:2010ji,Krumhansl1982}.
The profile of listeners' ratings of probe-tones reflects a tonal hierarchy, and it is this hierarchy of pitch stabilities that plays a large role in governing tonal perception. The extent to which different music listening behaviors and one's musical `culture' influence tonal perception is an open question, although evidence exists that Western classical music training results in differentiated, and often more nuanced, pitch expectations and probe-tone profiles \cite{cuddy1987recovery,bidelman2013tone,krumhansl1991music,tillmann2008music}.

To model these types of findings, computational models of tonal perception typically aim to provide methods that, given a musical context, compute a response that can be judged to be more or less appropriate for the implicit tonality of that context.
Given the predominance of the probe-tone paradigm for studies of human tonal perception, a common practice is to elicit a \emph{quasi}-goodness-of-fit response from the model for a probe-tone given a musical stimulus, such that the responses can be compared to human probe-tone ratings (e.g. \cite{leman95:_model_retroac_tone_center_percep,Tillmann:2000,Cancino14:ismir,large16:_neurod_accoun_music_tonal}).
Another way to judge the responses is to define a metric over the responses and compare the resulting topology to geometric constructs from music theory, such as the Tonnetz \cite{Toiviainen:2003ey}, a toroidal representation of key distance \cite{krumhansl82:_tracin}, or the circle of fifths \cite{Cancino14:ismir}. 

The computational models proposed in the literature tend to emphasize one of various different factors that are thought to play a role in tonal perception.
Whereas some works seek to explain empirical results mainly by a computational account of the lower levels of the auditory system~\cite{leman95:_model_retroac_tone_center_percep,large16:_neurod_accoun_music_tonal}, others focus more strongly on the role of long-term memory in tonal perception~\cite{LemanCarreras:96b,Tillmann:2000,Cancino14:ismir}.

Of the models that involve some representation of long-term memory, most do not account for that representation in an \textit{ecologically plausible} manner, meaning that there is no plausible simulation of how the long-term memory representations come about as a result of long-term exposure to music.
First, long-term memory is usually modeled by some form of self-organization of static representations of musical contexts or events, producing a low-dimensional map of musical stimuli, in which the neighborhood relationship captures semantic information (such as tonal affinity) \cite{Toiviainen:2003ey,LemanCarreras:96b,Tillmann:2000,Cancino14:ismir}.
Although the principle of self-organization has been used to account for the structure of cortical maps such as those in the visual cortex~\cite{durbin:dimension}, there is no evidence that this principle also underpins long-term memory.
Moreover, the fact that musical contexts are being mapped as static entities is at odds with the fundamentally temporal nature of the music listening process. 
As formalized in the \emph{predictive coding} framework~\cite{friston05}, an increasingly prominent idea in cognitive science is that of anticipation as a universal driving force for cognition~\cite{Dennett1991-DENCE,Clark2013-CLAWNP}.



Music researchers have also focused on the temporal dynamics of tonal and harmonic expectations (e.g., \cite{sears2014perceiving} and \cite{schmuckler:1994}), and some models based on self-organizing maps (SOMs) \cite{koho82a}
  do account for effects of temporal order in musical listening \cite{leman95:_model_retroac_tone_center_percep, Tillmann:2000}. 
A limitation, however, is that these effects are not taken into account in the training of the maps, representing the learning process that forms long term memory of music. 
Toiviainen and Krumhansl \cite{Toiviainen:2003ey} also employ a SOM, but, as they state, use it for visualization purposes, and not ``to simulate any kind of perceptual learning that would occur in listeners through, for
instance, the extraction of regularities present in Western music.''
Although the model offered by \cite{collins2014combined} does learn from musical sequences to predict tonal expectation in listeners, the model itself does not use sequential tonal information to learn and drive its predictions.

Another limitation of most long-term memory models for tonal learning is that they work with stimuli that are reduced in one or more ways.
For example, the input may consist of discrete representations of tones such as MIDI note numbers~\cite{Cancino14:ismir}, pitch classes~\cite{Tillmann:2000}, artificial harmonic representations~\cite{agres15:_co}, or of artificial harmonic sounds such as Shepard tones~\cite{leman95:_model_retroac_tone_center_percep}. 
Furthermore, the musical material that a model is exposed to may be limited to monophonic melodic lines~\cite{Cancino14:ismir}, sets of chords or harmonic cadences~\cite{leman95:_model_retroac_tone_center_percep}, or even a set of probe-tone profiles~\cite{Toiviainen:2003ey}. A notable exception to this is~\cite{LemanCarreras:96b}, which uses an audio recording of Bach's Well Tempered Clavier (WTC), performed on a harpsichord, to train a SOM by converting the acoustic signal to \emph{auditory images}.
The work of \cite{collins2014combined} also takes an ecological approach by using real audio and plausible psychological representations, with multiple representations along the sensory-cognitive spectrum, to better account for human tonal expectation.



The central question of this work is whether sequential predictive models of musical memory induce memory representations that convey tonal structure, similar to the static self-organizing models that are predominant in computational modeling of tonal learning.
To answer this question, we employ Recurrent Neural Networks (RNNs) and variants such as Long Short Term Memory (LSTM)~\cite{Hochreiter:1997ks}, which provide a common and effective modeling approach to the task of predicting future input from a history of past inputs.
A further objective is to see whether tonal expectations can also be elicited in the models by training on ecologically valid musical data rather than artificial data.
The present work approaches ecological validity in four ways: 1) using commercial audio recordings rather than symbolic or reduced music, 2) employing a psychoacoustically plausible input representation (the Constant-Q representation), 3) training corpora that span more than one genre (Bach and the Beatles) to better reflect a lister's musical experience, and 4) using more than 1 key to train the model (much related research transposes the training dataset to one key, e.g. \cite{Cancino14:ismir,Agres:2015vm}).
We test the effect of the training data on the strength and character of the tonal expectations of the model. Furthermore we measure the impact of shuffling the training data to gauge the importance of the sequential order of the music. Finally, we investigate the relationship between the training objective of the models (to predict the immediate future based on the present and past), and the strength of tonal hierarchy in the model expectations.


The paper is structured as follows. 
In Section~\ref{sec:method}, we provide a brief description of both the audio representation and of the predictive RNN models used in our experiments.
Section~\ref{sec:experiments} briefly reviews the datasets used to train the RNN models, and presents and discusses a comparison of model predictions to the results of probe-tone experiments.
Finally, conclusions and future work are presented in Section~\ref{sec:conclusion}.

\section{Method}\label{sec:method}
In this Section we describe the predictive models we use for our experiment (Section~\ref{sec:models}), and the audio representation used to present the data to the models (Section~\ref{sec:constant-q-transform}).

\subsection{Constant-Q Transform}\label{sec:constant-q-transform}
The Constant-Q Transform (CQT)~\cite{Brown:1991gl} is a discrete frequency domain representation of audio.
Although the CQT was not conceived explicitly as a model of the human auditory periphery, it shares an important characteristic with such models in that it samples the frequency axis logarithmically---a psychoacoustically plausible feature, since human listeners tend to perceive pairs of tones as equidistant when their respective frequency \emph{ratios} are equal.
The CQT is widely used in applications involving musical audio, since its frequency bins can be configured to match the 12 tone octave division of Western music.
To obtain a CQT spectrogram, conveying the change in frequency content of audio over time, the CQT can be computed over series of consecutive short, windowed segments of the audio, analogous to the Short-Time Fourier Transform. 

\subsection{Recurrent Neural Networks}\label{sec:models}
An RNN is a neural architecture that allows for modeling dynamical systems \cite{Graves:2013sm}. 
Let $\vect{x}_1, \dots, \vect{x}_t$ be a sequence of $N$-dimensional (normalized) input vectors and $\vect{y}_1, \dots, \vect{y}_t$ be its corresponding sequence of outputs.
An RNN provides a natural way to model $\vect{x}_{t+1}$, the next event in the sequence, by using the outputs of the network to parametrize  a predictive distribution given by
\begin{equation}
p(x_{t+1,i} \mid \vect{x}_{t}, \dots, \vect{x}_1) = y_{t, i}
\end{equation}
where $x_{t+1, i}$ and $y_{t,i}$ are the $i$-th component of $\vect{x}_{t+1}$ and $\vect{y}_{t}$ respectively.

The basic component of an RNN is the \emph{recurrent layer}, whose activation at time $t$ depends on both the input at time $t$ and its activation at time $t-1$.
Although theoretically very powerful, in practice RNNs with \emph{vanilla} recurrent layers are known to have problems learning long term dependencies due to a number of problems, including vanishing and exploding gradients \cite{Pascanu:2013tw}.
Other recurrent layers such as LSTM layers \cite{Hochreiter:1997ks} and gated recurrent units (GRUs)\cite{chung2014empirical}  try to address  some of these problems by introducing special structures within the layer,  such as purpose-built memory cells and gates to better store information.
More recently, recurrent layers with multiplicative integration  (MI-RNNs) \cite{Wu:2016vm} have been shown to extend the expressivity of traditional additive RNNs by changing the way the information from different sources is aggregated within the layer while introducing just a small number of extra parameters.

Given a training set consisting of inputs and targets, the parameters of an RNN can be learned in a supervised fashion by minimizing the cross entropy ($\mathit{CE}$) between its predictions and the targets.

A more thorough description of RNNs lies outside of the scope of this paper. 
For a more mathematical formulation of LSTMs and GRUs, we refer the reader to \cite{Graves:2013sm,chung2014empirical}. 
A more detailed description of MI-RNNs can be found in the Appendix of \cite{Wu:2016vm}.

\section{Experiments}\label{sec:experiments}
In this Section we describe the two datasets used for the experiments in this paper (Section \ref{sec:datasets}) and briefly review the theoretical framework of probe-tone experiments (Section \ref{sec:probetone-exp}), as well as a description of the training procedure (Section \ref{sec:training}). In Section \ref{sec:results-discussion} the results of the probe-tone experiments are presented and discussed.
\subsection{Probe-tone experiments}\label{sec:probetone-exp}

A probe-tone test is an experimental framework  to quantitatively assess the hierarchy of tonal stability \cite{krumhansl1990}.
This experimental framework consists of a set of musical stimuli like scales or cadences that unambiguously instantiate a specific musical context, such as a key. 
After presenting the stimulus, a participant hears a set of probe-tones, usually the set of 12 pitch classes, and the participant, either a human participant or a computer model, is asked to rate on quantitatively how well the probe-tones fit the musical stimulus.

Let $\mathbf{X} = \{\vect{x}_1, \cdots \vect{x}_T\}$ be an input musical stimulus, and 
$\mathbf{T} = \{\vect{\tau}_1, \dots, \vect{\tau}_{12}\}$ the set of probe-tones each corresponding to one of the 12 pitch classes. 
In order to quantitatively assess how well a probe-tone $\vect\tau$ fits the musical stimulus, we compare $\vect{y}^*$, the predictions of the RNN given the input stimulus, and the probe-tone using the Kullback-Leibler (KL) divergence.

In this paper, we use the above described model to reproduce the classic Krumhansl and Kessler (KK) probe-tone experiment \cite{krumhansl82:_tracin}.
This study is interesting for us mainly because
1) the probe tone contexts are polyphonic, featuring scales, chords, and cadences, thus highlighting capability of the proposed model to process polyphonic data, and
2) only expert listeners were tested (the participants of this experiment had an average of 11 years of formal music education), allowing us to directly compare the expectations of the model to those of an expert listener.
The setup for this  experiment requires a set of 14 tonal contexts\footnote{See Table 1 in \cite{krumhansl82:_tracin}.}:  ascending major and (harmonic) minor scales, three chord cadences (II-V-I, IV-V-I, VI-V-I) in both major and minor and individual chords (major triad, minor triad, dominant seventh chord and diminished chord).
In our experiments, we transpose each context to every key, yielding 12 variants of each context.
In order to aggregate the results over all keys, we average the KL divergence for each context.

Following the original experimental setup, both stimuli and probe-tones are generated using Shepard-tones, which consists of five sine wave components in a five-octave range from 77.8 Hz to 2349 Hz, with an amplitude envelope such that the low and high ends of the range approached hearing threshold \cite{krumhansl1990}.

We use Pearson's correlation coefficient to  compare the goodness-of-fit of the probe-tones learned by the models with the KK probe-tone ratings. 

\subsection{Datasets}\label{sec:datasets}
The WTC is a collection of 96 pieces for solo keyboard, consisting of two sets of 24 Preludes and Fugues in each key.
Composed by Johann Sebastian Bach, the WTC is widely recognized as one of the most important works in Western music.  
We use a performance of the WTC by renowned Canadian pianist Angela Hewitt\footnote{Hyperion CDS44291/4 1998}. The total duration of this recording is 4.5 hours.
We perform data augmentation on the WTC dataset by pitch shifting each recording between $-6$ and $+5$ semitones using \emph{pyrubberband}\footnote{\smallurl{https://github.com/bmcfee/pyrubberband} Accessed April 2017.}.
We thus obtain 1152 pieces for the WTC, equivalent to nearly 53 hours of music.

Additionally, we use a second dataset consisting of 12 Albums by The Beatles, with a total of 179 songs with an approximate duration of 7.5 hours. We do not perform data augmentation on the Beatles data\footnote{
Exploratory experiments showed that using pitch shifting on the Beatles songs worsened the predictions of the RNNs.
This worsening might be due to the fact that most of these recordings include several instruments and voices, including unpitched percussion instruments.}.

To facilitate the exposure of the models to regularities in the change of pitch content over time, we do not compute the CQT spectrograms by taking equidistant frames in absolute time, but instead link the spectrogram frame rate to the musical time, such that the instantaneous frame rate is always an integer multiple or submultiple of the beat rate.
For the Beatles data, we do so by using publicly available beat annotations\footnote{\smallurl{http://isophonics.net/content/reference-annotations-beatles}. Accessed April 2017.}.
For the WTC recording by Hewitt no such annotations were available, but versions in Humdrum format of the pieces were obtained from KernScores\footnote{\smallurl{http://kern.ccarh.org}. Accessed April 2017.}. 
The Humdrum files were converted into MIDI files, which were manually edited using MuseScore to match the repetitions as performed by Hewitt.
By aligning piano-synthesized audio renderings of the MIDI files to the Hewitt recordings using the method described in~\cite{grachten13:_autom}, beat times were automatically inferred for the recordings.

Based on the typical temporal densities of musical events in the two datasets, we chose a temporal resolution of a quarter beat for the CQT spectrogram in the case of the Beatles, and a sixteenth beat in the case of WTC.
We will return to this issue in Section~\ref{sec:bias-learn-towards}.

Each slice of the CQT spectrogram is a 334-dimensional vector that represents frequencies between 27.5 and 16744.04 Hz with a resolution of 36 frequency bins per octave.
This configuration was chosen to avoid spectral leakage between adjacent frequency bins, and is similar to the one used by Purwins et. al.  \cite{purwins2000new}.
Additionally, this configuration is also able to accommodate at least the fundamental frequency plus at least three harmonics to the highest note of a piano. 
We normalize each slice of the CQT to lie between 0 and 1.

\subsection{Training}\label{sec:training}
For the experiments in this paper we use RNNs as described in Section~\ref{sec:models} as a sequential alternative to the static models typically used for tonal learning, such as SOMs and RBMs.
To get an impression of the performance of sequential models in general for this task, we test five different variants of the recurrent layer, namely a vanilla RNN (vRNN), an LSTM, a GRU, and two models with multiplicative integration: a vanilla recurrent layer (vRNN/MI) and an LSTM/MI\footnote{In the current experiments the GRU/MI yielded pathological results, possibly due to an implementation problem.}.
In all variants, the model has a single hidden recurrent layer with 75 $\tanh$ units and an output layer with sigmoid units.
The use of different model variants also allows us to investigate the relationship between the prediction error and the similarity of model expectations to human goodness-of-fit ratings of probe-tones.

In order to investigate the kind of statistical regularities in music that produce human-like probe-tone results, we train each model on two different versions of each dataset, namely training the model using the original data, and training the  model shuffling the spectrograms in a piece-wise fashion.
Randomizing inputs per piece preserves the global pitch distribution of the piece but disrupts temporal cues to musical expectations, like harmonic progressions and voice-leading.


We split each dataset into 5 equally sized non-overlapping folds, resulting in $4$ RNN architectures $\times$ $2$ orderings of the CQT spectrograms (original vs. randomized spectrograms) $\times$ $5$ folds $\times$ $2$ datasets $=$ $80$ trained models.
For each fold, 80\% of the pieces (ca.~184  pieces for the WTC and 29 for the Beatles) are randomly selected to be used for training and 20\% for testing (ca.~46 pieces for the WTC and 7 for the Beatles).
The predictive accuracy of each model is measured by the mean cross-entropy (MCE) on the test set.
 The models are trained using \mbox{RMSProp}~\cite{Tieleman:2012kl}, a variant of stochastic gradient descent that adaptively updates the step-size using a moving average of the of the magnitude of the gradients. 
The initial learning rate is set to $10^{-3}$.
The gradients are computed using truncated back propagation through time, where computation of the  gradients is truncated after 100 steps and are clipped at 1.
Each training batch consists of 20 sequences of 100 CQT slices. 
Each sequence is selected randomly out of the training data.
Thus, an epoch of training corresponds to the model seeing roughly the same number of time steps as in the whole fold.
Early stopping is used after 100 epochs without any improvement in the test set.
All RNNs are implemented using \emph{Lasagne}\footnote{\smallurl{https://github.com/Lasagne/Lasagne}. Accessed April 2017.}.
We provide online supplementary materials describing all of the technical details for performing the probe-tone experiments in this paper\footnote{\smallurl{http://carloscancinochacon.com/documents/online_extras/ismir2017/sup_materials.html}.}.



\subsubsection{Biasing learning towards predicting change}\label{sec:bias-learn-towards}
A crucial question when applying discrete time recurrent models to a continuous stream of data such as audio is how to choose the rate of discrete time steps with respect to the absolute time of the data.
This choice depends on the approximate rate or temporal density of relevant events in the data---in our case the notes that make up the musical material.
Ideally, we would like the discrete time steps to be small enough to capture the occurrence of even the shortest notes individually, but if the discrete time step is chosen much smaller than the median event rate, this leads to strong correlations between data at consecutive time steps.
A result of this is that training models to predict the data at time step $t+1$ teaches them to strongly expect the data at $t+1$ to be approximately equal to the data at $t$.
Choosing a larger discrete step size for the model alleviates this problem, but has the disadvantage that the data the model sees at a particular time may actually be an average over consecutive events that happened within that larger step.

We slightly revise the training objective of the models as a remedy to this unfortunate trade-off.
This revised objective biases the models to care more about correctly predicting the data at $t+1$ when the change from $t$ to $t+1$ is large (e.g. the start of a new note) than when it is small (e.g a transition without any starting or ending note events).
This allows us to use a relatively small step size without causing the models to trivially learn to expect the data to stay constant between consecutive time steps.

More specifically, we modify the original cross-entropy objective $\mathit{CE}_{t}$ 
by multiplying it with a time-varying weight $w_{t}$ as follows:
\begin{equation}
\tilde{\mathit{CE}}_{t} \leftarrow w_{t} \mathit{CE}_{t}, 
\end{equation}
where $w_{t}$ is given by
\begin{equation}
w_{t} = \left\{\begin{array}{ll}1 & \mbox{if $\sum_i^{N}|x_{t + 1, i} - x_{t, i}|> \varepsilon$} \\ \beta & \mbox{otherwise}\end{array}\right.
\end{equation}
where $\varepsilon \in\mathbb{R}$ acts as a threshold distinguishing small and large change transitions, and $\beta \in\mathbb{R}$ controls the relative influence of prediction errors on the training in the case of small change transitions\footnote{We empirically found a binary distinction between small and large change transitions to be more effective than a gradual weighting scheme}.
Based on an informal inspection of the model predictions in a grid search on $\beta$ and $\varepsilon$, we choose $\beta = 10^{-3}$, and $\varepsilon$ such that

\begin{equation}
P_{\mathit{training}}(\sum_i^{N}\ \lvert x_{t + 1, i} - x_{t, i} \rvert\ \le\ \varepsilon) = 0.505   
\end{equation}

where $P_{\mathit{training}}(X)$ denotes the empirical probability of event $X$ under the training data. 


\subsection{Results and Discussion}\label{sec:results-discussion}
\begin{figure}[t]
  \centering
  \includegraphics[width=.5\textwidth]{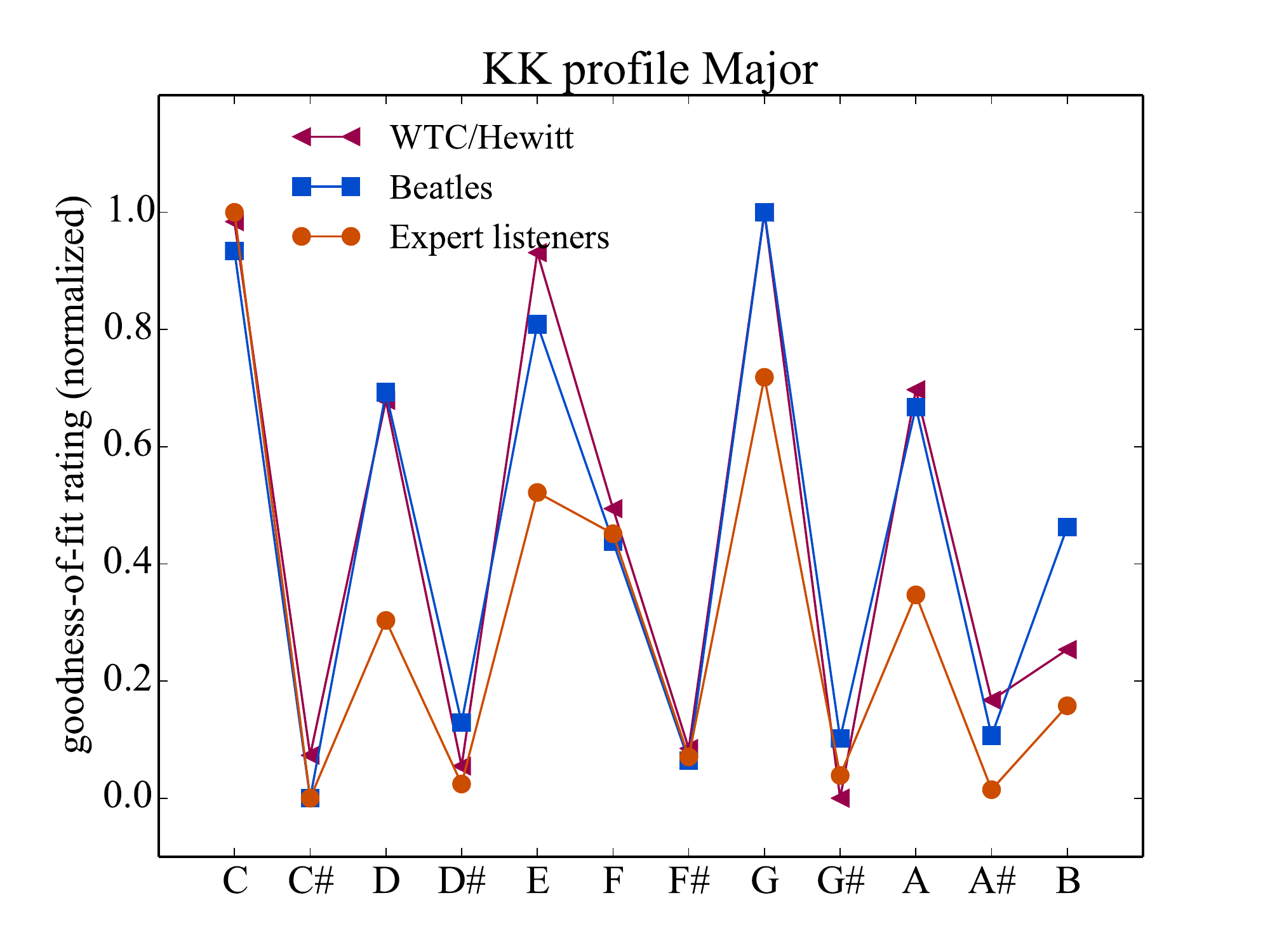}

  \includegraphics[width=.5\textwidth]{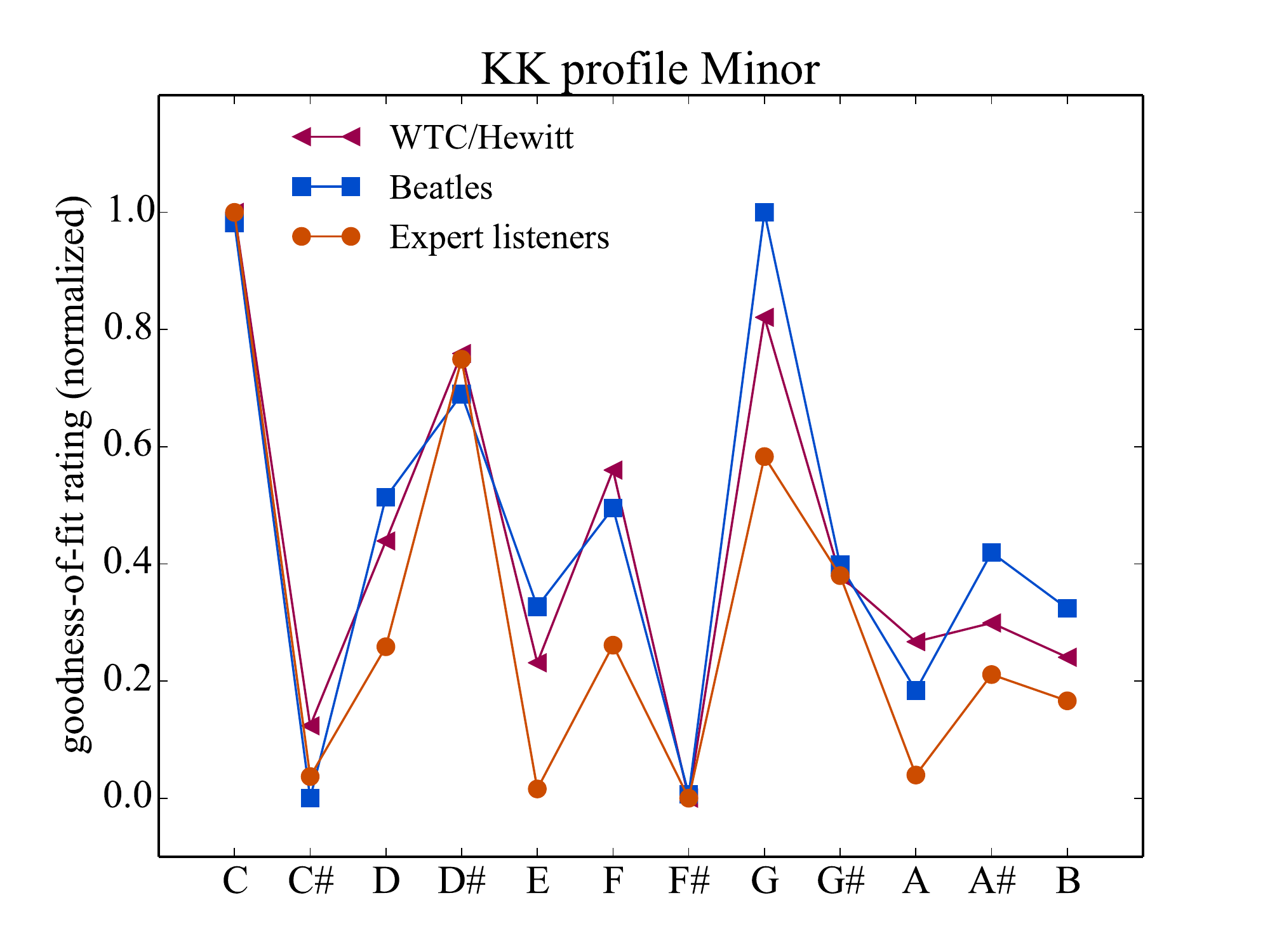}
 
  \caption{Expectations of the models trained on WTC and Beatles datasets compared to average probe-tone ratings by expert listeners for major and minor contexts~\cite{krumhansl82:_tracin}}
  \label{fig:kk_minor_major}
\end{figure}

Figure \ref{fig:kk_minor_major} compares the aggregation of the probe-tone ratings (see Section~\ref{sec:probetone-exp}) for both major and minor contexts with the expectations of the best predictive models (as in lowest MCE in the test set) for each dataset, which in both cases is the GRU trained without shuffling the data.
Table \ref{tab:kk_corr} shows the correlation between the KK profiles and the model expectations.
All of the correlations are statistically significant ($p<0.0002$).
Although the values obtained for the models trained on the Beatles data are slightly lower, the strength of the correlations between the empirical data and the model simulation is on a par with those reported in the literature \cite{LemanCarreras:96b,Tillmann:2000}.
Pairwise two-sample Kolmogorov--Smirnov tests (KK vs. Hewitt/WTC, KK vs. Beatles and WTC/Hewitt vs. Beatles) reveal that the three profiles are not significantly different from one another ($p\geq 0.19$).

\begin{table}[!h]
\centering
\caption{Pearson's correlation between normalized predictions of the model with the lowest mean cross-entropy for each dataset and KK major and minor profiles}
\vskip1ex
\small

\begin{tabular}{ccc}
\toprule
          & KK major & KK minor \\ \midrule
WTC/Hewitt & 0.915    & 0.940    \\
Beatles    & 0.900    & 0.885    \\ \bottomrule
\end{tabular}
\label{tab:kk_corr}
\end{table}


  
\begin{figure}[t]
  \centering
  \includegraphics[width=.5\textwidth]{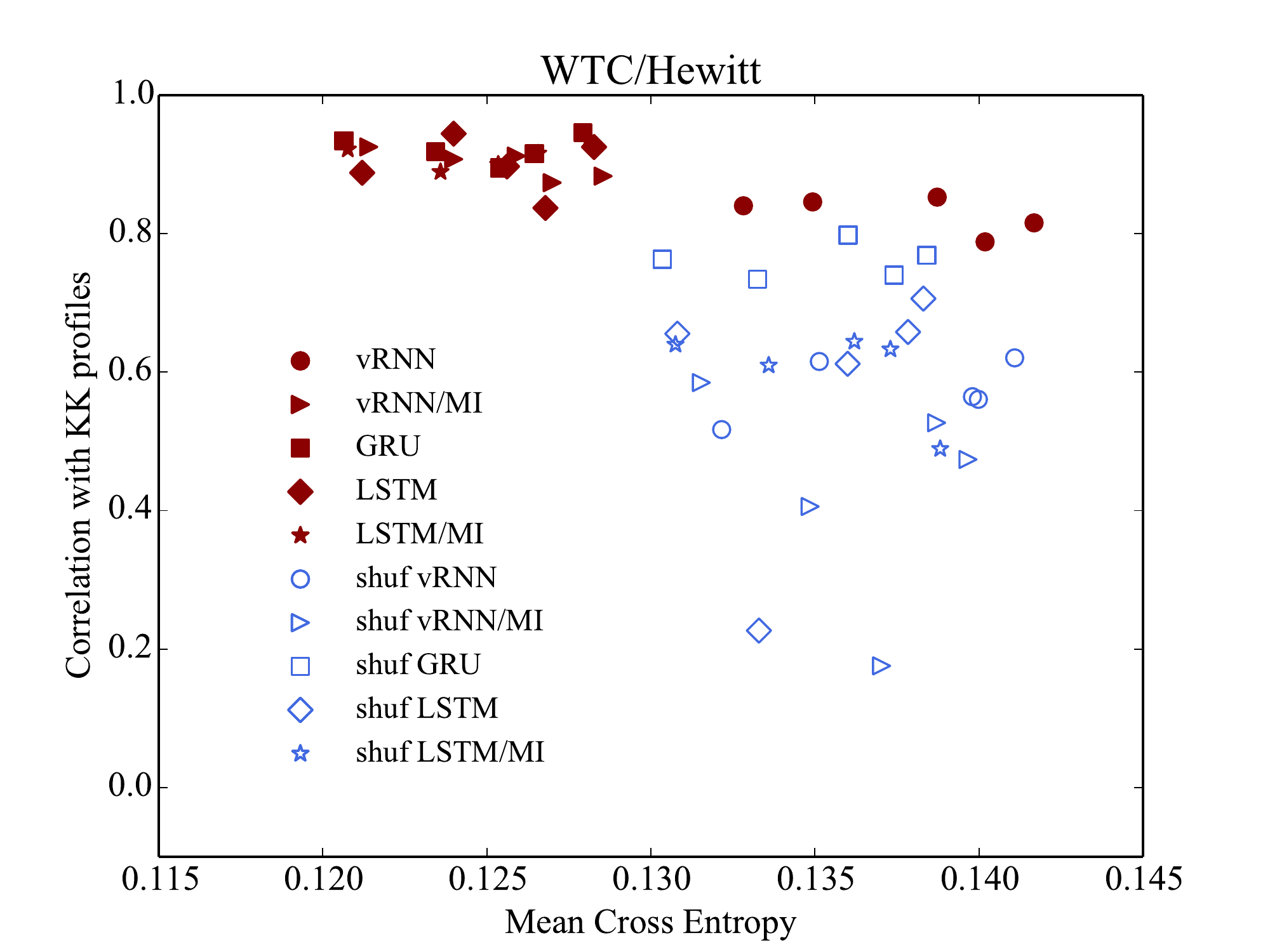}
  
  \includegraphics[width=.5\textwidth]{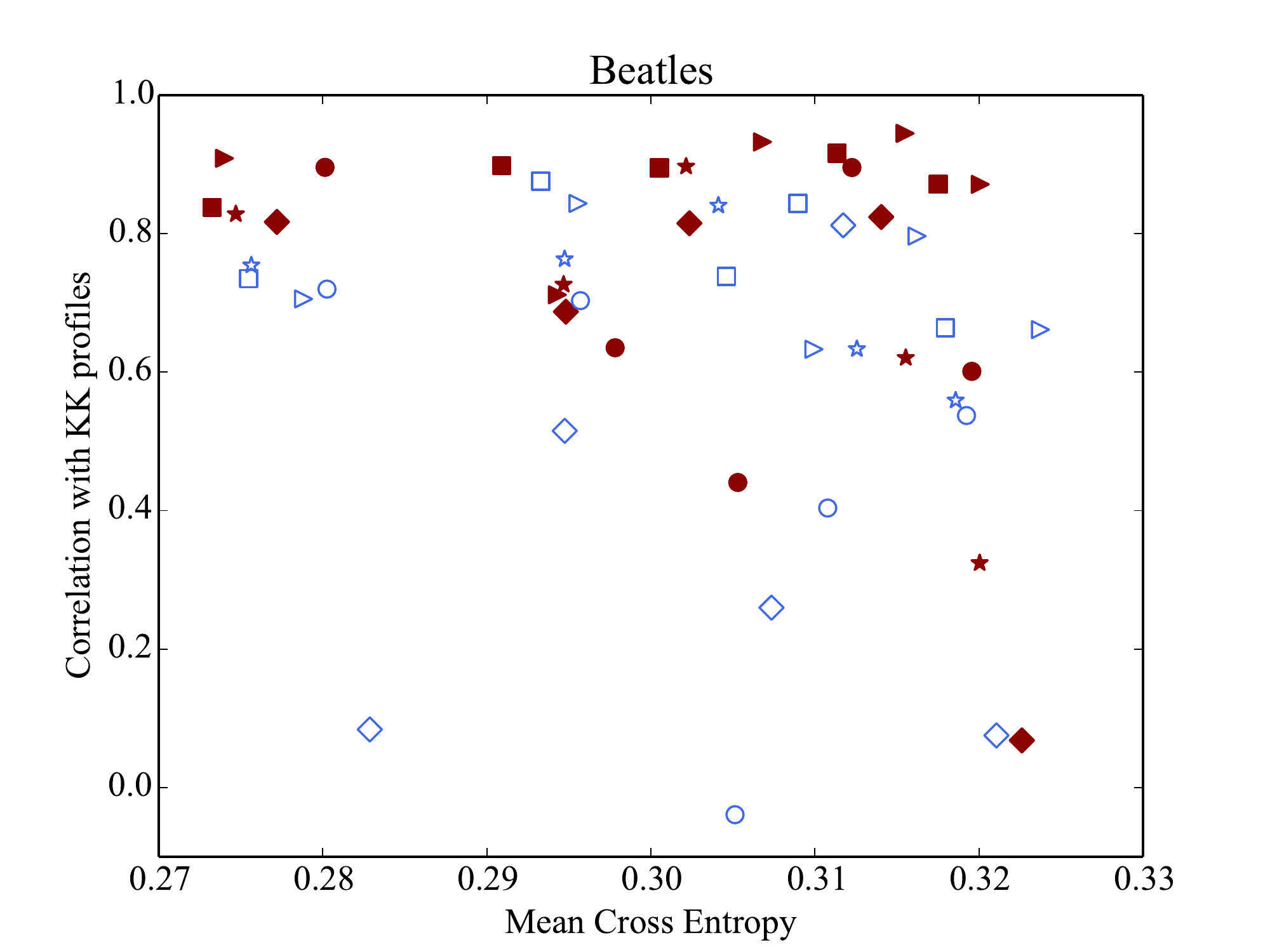}
  \caption{Similarity of model expectations to human probe-tone ratings (Pearson correlation coefficient) plotted versus
    the mean cross-entropy of the models over a test set; \emph{shuf} denotes models trained on shuffled data}
  \label{fig:mce-corr}
\end{figure}

The above result shows that the expectations of the proposed models reflect the tonal characteristics of the musical context that evoked those expectations.
This is expected but not trivial, since the training objective of the models is solely to predict how a given sequence of musical information (in the form of CQT spectrograms) will continue.
An interesting question is therefore whether there is any relation between the predictive accuracy of a model (that is, how successfully it predicts future musical events based on the music up to now), and the correlation of its probe-tone response to that of human subjects.
In the plots of Figure~\ref{fig:mce-corr}, the vertical axis measures the Pearson correlation coefficient of the probe-tone responses of different models with the KK profiles, and the horizontal axis measures predictive accuracy of the models, in terms of their MCE over the test data.
For each model type in the legend, there are five different scatter points, representing models trained on each of five non-overlapping folds of the data (see Section~\ref{sec:training}).
The vertical coordinate of each scatter point is the result of averaging the correlation coefficients of responses to all transpositions of the probe-tone stimuli (see Section~\ref{sec:probetone-exp}).

The scatterplots for the WTC and Beatles show that on average, MCE is higher for models trained the Beatles data than for those trained on WTC.
This is likely due to the fact that the WTC data are single instrument recordings (piano) with relatively homogeneous CQT spectrograms, whereas the Beatles recordings are multi-instrumental, leading to more dense and complex CQT spectrograms.

For the WTC data, training models on shuffled CQT data has a noticeable negative impact on both predictive accuracy and tonal expectations. For the Beatles data this effect is less pronounced.
There may be multiple explanations for this.
First, even if the WTC data, being solo piano recordings, are spectrally simpler, they are probably more complex both harmonically and melodically than the Beatles data.
As such, shuffling the data temporally is more of a disruption to the WTC data than to the Beatles data.
Secondly, the WTC pieces tend to include brief modulations away from the main key of the piece.
This means that shuffling the data within a piece may mix data from different keys, making prediction more dificcult.

Despite these differences, both WTC- and Beatles-trained models roughly show the same overall pattern: models with low predictive error have high KK correlations, whereas models with high predictive error may or may not have high KK correlations.
This suggests that in order to form accurate musical expectations, it is indispensable to have a notion of tonal structure.
But conversely, having a notion of tonal structure by itself is not a sufficient condition for accurate musical expectations.
This implies that there are other factors beyond tonality, such as voice leading, rhythm, and cadential structure, that help predict how a given musical context will 
continue (see \cite{schmuckler1994harmonic} and \cite{sears2014perceiving} for behavioral evidence to this effect).

%
%
%
%

\section{Conclusion}\label{sec:conclusion}
In this paper we showed that the expectations of ecologically trained predictive models of music exhibit tonal structure very similar to that observed in humans through probe-tone experiments.
We believe this finding is relevant, since most computational modeling approaches to tonal perception that involve a representation of statistical regularities in musical data do not account for the perceptual learning of such regularities in a plausible way.
The musical expectations of the models used here are formed by training the model to reduce the prediction error for future musical events based on the musical context up to the present---a cognitively plausible task according to the predictive coding theory of the brain \cite{Clark2013-CLAWNP}.
Furthermore, we demonstrate that tonal learning within such models is not only possible based on training data known to exhibit rich tonal qualities (Bach's WTC, artificial cadences), but also occurs as an effect of exposure to audio representations of ``real-world'' popular and harmonically simpler music (The Beatles). 
This more accurately mirrors the kind of musical exposure people have, even if real-world musical enculturation would typically involve a wider range of music.

An analysis of the relation between the predictive accuracy of the model and the degree of tonal structure exhibited by model expectations shows that tonal expectations are a necessary but not a sufficient condition for accurate musical expectations.
This suggests that there are other---presumably temporal---cues to musical expectation beyond tonal structure.
Evidence for this is the fact that models trained on temporally shuffled WTC data form less accurate expectations than models trained on the ordered data.
This effect is not observed for the Beatles data, possibly because of its simpler melodic and harmonic structure.

The empirical validation of the models we presented here offers various further avenues of research that we have not yet pursued.
For example, a qualitative analysis of the learned representations of the models may provide further insights into the cues that influence musical expectations.
In models with multiple hidden layers, an interesting question is where the different learned representations lie along the sensory-cognitive spectrum of tonal representations, as hypothesized by \cite{collins2014combined}.

  \newpage
  \section{Acknowledgments}
  This work has been partly funded by the European Research Council (ERC) under the EU’s Horizon 2020 Framework Programme (ERC Grant Agreement No. 670035, project CON ESPRESSIONE). We thank Carol L. Krumhansl for providing the probe-tone data.

\IfFileExists{bibliographies/bib_mg.bib}{
\bibliography{bibliographies/bib_mg,bib_cc}
}{
\bibliography{bib_cc,bib_mg}
}

\end{document}